\documentclass[preprint,12pt]{elsarticle}

\usepackage{color}
\usepackage{bm}
\usepackage{amsmath}
\usepackage[hang,small,bf]{caption}
\usepackage[subrefformat=parens]{subcaption}
\usepackage{graphicx}

\usepackage{amssymb}

\begin{document}

\begin{frontmatter}



\title{
Long-Term Modeling of Financial Machine Learning\\
for Active Portfolio Management
}


\author[label1]{Kazuki Amagai}
\author[label1,label2]{Tomoya Suzuki}

\affiliation[label1]{organization={Graduate School of Science and Engineering, Ibaraki University},
            addressline={4-12-1, Nakanarusawa}, 
            city={Hitachi},
            postcode={316-8511}, 
            state={Ibaraki},
            country={Japan}}

\affiliation[label2]{organization={Daiwa Asset Management Co.Ltd.},
            addressline={1-9-1, Marunouchi}, 
            city={Chiyoda-ku},
            postcode={100-6701}, 
            state={Tokyo},
            country={Japan}}

\begin{abstract}
In the practical business of asset management by investment trusts and the like, the general practice is to manage over the medium to long term
owing to the burden of operations and increase in transaction costs with the increase in turnover ratio. 
However, when machine learning is used to construct a management model, the number of learning data decreases with the increase in the long-term time scale; 
this causes a decline in the learning precision. 
Accordingly, in this study, data augmentation was applied by the combined use of not only the time scales of the target tasks but also the learning data of shorter term time scales,
demonstrating that degradation of the generalization performance can be inhibited even if the target tasks of machine learning have long-term time scales. 
Moreover, as an illustration of how this data augmentation can be applied, 
we conducted portfolio management in which machine learning of a multifactor model was done by an autoencoder and mispricing was used from the estimated theoretical values. 
The effectiveness could be confirmed in not only the stock market but also the FX market, and a general-purpose management model could be constructed in various financial markets.

\section*{Highlights}
\begin{itemize}
\item Long-term modeling is necessary for practical use of financial machine learning.
\item Data augmentation with multi-time scale data is useful for long-term modeling.
\item Multi-factor model can be improved by the autoencoder using data augmentation.
\item Our model can be applied to portfolio management in various financial markets.
\end{itemize}
\end{abstract}

\begin{keyword}
machine learning \sep data augmentation \sep anomaly detection \sep financial timeseries analysis \sep econophysics
\PACS 07.05.Kf \sep 07.05.Mh \sep 89.65.Gh \sep 89.75.-k
\end{keyword}

\end{frontmatter}







\section{Introduction}
In asset management businesses, such as portfolio management, 
it is common to operate in the medium to long term due to the increase in operational burden and transaction costs.
However, to compose a longer-term model
the number of usable learning data decreases; 
hence, the model performance declines. 
Accordingly, in this study, the concept of data augmentation\cite{Aug1,Aug2},
which has chiefly been utilized in the fields of image recognition and natural language processing,
was applied to longer-term modeling for financial machine learning. 
As a practical example, 
a multifactor model\cite{MF1,MF2} is learned using data augmentation for portfolio management,
where we use not only the main timescale data for rebalancing the portfolio
but also shorter timescale data (i.e., using multi-timescale data) to enlarge the number of learning data for machine learning.
%

The multifactor model is a conventional asset price model in which the presence of common latent variables called factors is assumed,
and the process of price formation of various financial assets is modeled. 
In particular, the Fama-French and BARRA type multifactor
models, which assume conventional assets, such as stocks and bonds, are the main form thereof; 
however, the identification and verification of the presence of factors is progressing in non-conventional assets, such as virtual currencies (crypto-assets) as well\cite{cripto}. 
In addition the presence of various factors has been reported in foreign exchange markets\cite{FX1,FX2,FX3,FX_factor}. 
However, the number of factors that are assumed to be effective has been increasing annually, and this has been questioned, such as in the case of the ``factor zoo,''
wherein distinguishing between useful, useless, and redundant factors is difficult\cite{Zoo}.
The information of most factors is probably overlapped,
and more substantive factors may be intrinsic. 
Moreover, even assuming that several substantive factors can be identified, there is a possibility that the effectiveness of each factor might change dynamically depending on the economic environment or phase of the business cycle. 
Accordingly, in this study, factors are extracted from the middle layer of autoencoder\cite{AE} in a data-driven manner
and the dynamic changes of the effectiveness of factors are dealt with by sequentially relearning the autoencoder.
Since the autoencoder is based on a neural network, it can express the nonlinearity and interactions of factors.

At the output layer of the autoencoder, the input information (return of the individual stocks) is restored by taking the linear combination of the non-linear factors that are extracted in the middle layer. 
This structure corresponds to the aforementioned multifactor model, and by the minimization of the restoration error, the restored output value corresponds to the theoretical value.
By comparing this with a realized value, the abnormality of a realized value is evaluated\cite{Gotou, Tsukahara}. 
This abnormal value is generally called a specific return, and not only separate circumstances (such as current news) that cannot be expressed by the common factors,
but also mispricing due to crowd psychology like overreaction, are assumed to be included therein. 
If mispricing is corrected to a fair value by the efficiency of the market, the abnormal value may have time series characteristics (momentum if it is an underreaction\cite{RM} or reversal if it is an overreaction\cite{RR}). 
In particular, a management strategy, like a specific-return-reversal strategy\cite{Barra}, also exists for a reversal. 
Accordingly, in the present study, we conducted portfolio management that uses the time series characteristics of abnormal values.

As extraction of abnormal values is crucial in this management policy, the learning performance of the autoencoder that estimates the theoretical values must be improved to the maximum possible extent. 
Accordingly, because the multifactor model built with the autoencoder does not depend on the characteristic time scale thereof, data augmentation that uses multiple time scales in combination may be a good fit. 
Moreover, because it does not depend on specific financial markets, it may be possible to build a general-purpose management model.
Accordingly, the present study constructed a quantile portfolio according to the deviation from the theoretical value and confirms the existence of pressure to correct mispricing,
with the goal thereof being substantiation of the effectiveness of data augmentation and portfolio management in not only the stock market but also the foreign exchange market.
As not only investor profits but also effects from correcting erroneous corporate assessments can be anticipated by managing portfolios
such that they conform to this correction pressure, this study aimed to construct a socially significant management model.

\section{Stock Selection for Portfolio Management}

\subsection{Multifactor model}
In this study, the time scale in portfolio management was set as $\tau$, and the rate of change of the stock price during that period (return) was $r$.
Moreover, if the stock price of company $i$ at date $t$ is $p_i(t)$,
the return on an individual stock can be written as follows:
\begin{equation}
	r_{i,\tau}(t)=\frac{p_i(t)-p_i(t-\tau)}{p_i(t-\tau)} \ .
	\label{return}
\end{equation}
This corresponds to a daily scale if $\tau=1$, and to a monthly scale if $\tau=20$. 
Five working days equate to one week in this article.

The multifactor model is a multiple regression model that expresses the return $r_{i,\tau}(t)$ on individual stocks by the linear combination of the common factors $F_{\tau,m}(t)$, such as size and value:
\begin{equation}
	r_{i,\tau}(t)=\sum_{m=1}^{M}\beta_{i,m}F_{\tau,m}(t)+\epsilon_{i,\tau}(t) \ .
	\label{factor_model}
\end{equation}
Here, $M$ is the number of factors, $\beta_{i,m}$ is the sensitivity to the $m$-th factor, and $\epsilon_{i,\tau}(t)$ is the residue. 
As the multifactor model can be defined by various time scales $\tau$ in this manner, it is not limited to a set time scale. 
Accordingly, in this study, the focus was on a multifactor model as one example in which data augmentation of multiple time scales may be effective.

\subsection{Utilization of autoencoder}
\begin{figure}[t]
	\centering
	\includegraphics[width=0.6\linewidth]{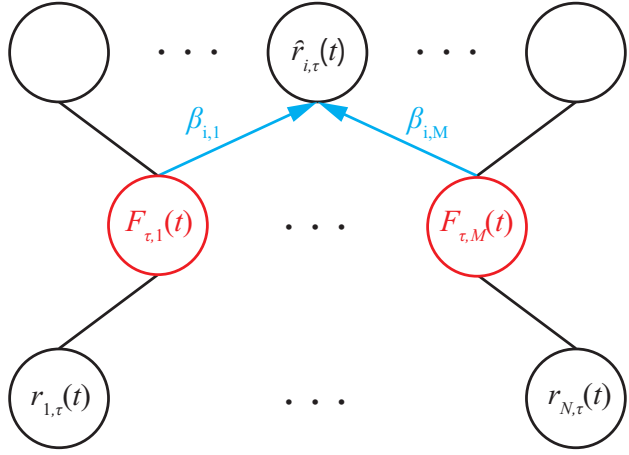}
	\caption{Structure of the autoencoder.}
	\label{fig:AE}
\end{figure}

An autoencoder can be utilized for flexible machine learning of a multifactor model\cite{MLFactor3}. 
In Equation (\ref{factor_model}), it is first necessary to identify the common factors $F_{\tau,m}(t)$ in advance, but $F_{\tau,m}(t)$ can also be simultaneously identified by the autoencoder. 
Accordingly, in this study, the multifactor model was built by an autoencoder (Figure \ref{fig:AE}).

The return vectors $\bm{r}_\tau(t)$ of all stocks ($i=1 \sim N$) at the same time are sequentially inputted to the input layer of the autoencoder:
\begin{equation}
	\bm{r}_\tau(t) = \{r_{1,\tau}(t), r_{2,\tau}(t), \cdots, r_{i,\tau}(t), \cdots, r_{N,\tau}(t) \} \ .
\end{equation}
Dimensional compression is performed at the middle layer, and the autoencoder is learned such that it can be restored at the output layer. 
If the learning can be done successfully, the output from the middle layer corresponds to the common factors $F_{\tau,m}(t)$,
and the combined weighting to the output layer corresponds to the sensitivity $\beta_{i,m}$.

\subsection{Utilization of mispricing}
When new information is inputted into a learned multifactor model, the theoretical value of the return can be estimated. 
Assuming that this estimated theoretical value is $\hat{r}_{i,\tau}(t)$, the following is obtained from Equation (\ref{factor_model}):
\begin{equation}
	r_{i,\tau}(t) = \hat{r}_{i,\tau}(t) + \epsilon_{i,\tau}(t) \ .
\end{equation}
Mispricing due to crowd psychology may also be included in this $\epsilon_{i,\tau}(t)$, in addition to the separate circumstances (specific return) of news, etc. that are not considered in the multifactor model.
\begin{equation}
	\epsilon_{i,\tau}(t) = \eta_{i,\tau}(t) + A_{i,\tau}(t) \ .
	\label{eq5}
\end{equation}
Here, $\eta_{i,\tau}(t)$ is the specific return and $A_{i,\tau}(t)$ is the mispricing.
$A_{i,\tau}(t)>0$ represents an overreaction, and $A_{i,\tau}(t)<0$ represents an underreaction.

If the specific return is a result that incorporates separate circumstances, it will not affect the subsequent stock price. 
However, in terms of mispricing, the correction pressure may exert an effect on the subsequent stock price due to the efficiency of the market\cite{Gotou}. 
Accordingly, by dividing all stocks into quantiles based on $\epsilon_{i,\tau}(t)$, a portfolio was constructed for which one can expect correction pressure in the minimum and maximum quantiles. 
Moreover, because the minimum quantile is an underreaction, the stock is bought (long), and because the maximum quantile is an overreaction, it is short sold (short), and management of a long-short portfolio is performed in this manner.

\section{Data Augmentation by Multiple Time Scales}
In the portfolio management described in the preceding section, the extraction of mispricing becomes crucial; 
therefore, the learning performance of the autoencoder must be improved to the maximum possible extent. 
Generally, in the case of machine learning, the greater the amount of learning data, the higher is the learning performance; 
hence, in this section, the learning data of the autoencoder are augmented by reproducing the return data of various time scales $\tau$ from the stock price time series $p_{i}(t)$ of the daily scale.

\begin{figure}[t]
	\centering
	\includegraphics[width=0.8\linewidth]{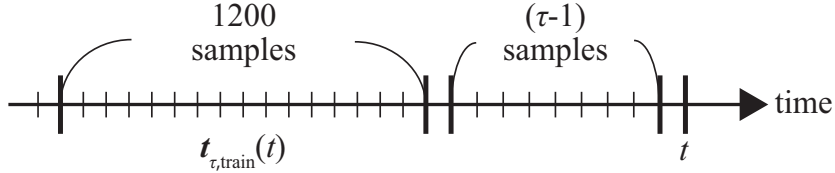}
	\caption{Schematic diagram of Equation (\ref{eq:trperiod}).}
	\label{fig:traindata2}
\end{figure}

Two pre-processings are performed. 
First, the number of learning data is standardized at 1,200 in all time scales to prevent bias from the time scales. 
Then, based on Equation (\ref{return}),
the return $r(t)$ at the time $t$ and within the closest $(\tau-1)$ contain the same price information $p(t)$;
thus, information leakage is prevented by excluding the times $t\!-\!1 \sim t\!-\!(\tau\!-\!1)$ from the learning targets(Figure\ref{fig:traindata2}).
Therefore, the learning time set $\bm{t}_{\tau, {\rm train}}(t)$ at time $t$ in the time scale $\tau$ becomes
\begin{equation}
	\bm{t}_{\tau, {\rm train}}(t)=\{ t\!-\!(\tau\!-\!1)\!-\!1, t\!-\!(\tau\!-\!1)\!-\!2, \cdots, t\!-\!(\tau\!-\!1)\!-\!1200 \} \ .
	\label{eq:trperiod}
\end{equation}
Second, because the scale of $r_{i,\tau}(t)$ differs in accordance with the time scale $\tau$, it is normalized by the standard deviation $\sigma_{\tau}(t)$ as follows:
\begin{eqnarray}
	\bm{r}^{\dag}_\tau(t) &=& \frac{\bm{r}_\tau(t)}{\sigma_\tau(t)} \label{eq7} \ ,\\
	\sigma_\tau(t) &=& S\{ r_{i, \tau}(t) \ | \ i=1 \sim N \} \ .
\end{eqnarray}
Here, $S\{ \cdot \}$ represents the standard deviation of the set.

From the above, if the time scale that serves as the learning target is set as $\tau^*$, the learning data set $\bm{S}_{\tau^*}(t)$ at time $t$ becomes
\begin{equation}
	\bm{S}_{\tau^*}(t) = \{ \bm{r}^{\dag}_\tau(t^\prime) \ | \ t^\prime \in \bm{t}_{\tau,{\rm train}}(t), \tau \in 1 \sim \tau^* \}\ .
	\label{eq:eqmts}
\end{equation}
In this manner, the data of multiple time scales is inputted into the autoencoder simultaneously and learning is performed.

When the autoencoder is learned by only a single time scale $\tau^*$, the learning data set $\bm{S}_{\tau^*}(t)$ becomes
\begin{equation}
	\bm{S}_{\tau^*}(t) = \{ \bm{r}^{\dag}_\tau(t^\prime) \ | \ t^\prime \in \bm{t}_{\tau,{\rm train}}(t), \tau \in \tau^* \}\ .
	\label{eq:eqsts}
\end{equation}

With the advance of time $t$, the autoencoder is relearned every time the learning data set $S$ is updated. 
The full combined type is used for each layer (input layer, middle layer, and output layer) of the autoencoder, while a linear function is used for the activation function of the output layer. 
Version 2.3.0 of tensorflow-gpu was used in the implementation of the autoencoder. 
Table \ref{tab:hyperparam} lists the various hyperparameters. 
The default values were employed for those items not listed in the table.

\begin{table}[t]
	\centering
	\caption{Hyperparameters of the autoencoder}
	\label{tab:hyperparam}
	\begin{tabular}{c|c}
		\hline
		Hyperparameter & Value used \\
		\hline
		epochs & 50 \\
		batch$\_$size & 128 \\
		optimizer & Adam \\
		Number of middle layers & 1 \\
		\hline
	\end{tabular}
\end{table}

\section{Analysis}

\subsection{Dataset}
In this study, performance management was conducted with the Japanese stock market as the object. 
As the management target stocks, the approximately 1,700 stocks that make up TOPIX were used. 
The period from January 2010 to December 2021 was used as the data period, the period from January 2016 to October 2017 was used
as the test period after model learning, and the period from January 2018 to December 2021 was used as the management period after model optimization. 
Stocks whose listing was canceled during the period were also used to avoid the survivorship bias, and the stock market at that time was faithfully reproduced and analyzed. 
All data were acquired from Refinitiv Eikon.

The time $t_{\rm test}$ of the test period advances by one time at a time, as shown below.
\begin{eqnarray}
	t_{\rm test} &\in& \bm{t}_{\tau, {\rm test}} \ ,\\
	\bm{t}_{\tau, {\rm test}}
	&=& \{ t_{\rm test}^{s}, t_{\rm test}^{s}\!+\!1, t_{\rm test}^{s}\!+\!2, \cdots, t_{\rm test}^{e} \} \ .
\end{eqnarray}
Here, $t_{\rm test}^{s}$ and $t_{\rm test}^{e}$ represent the start and end times of the test period, respectively. 
Meanwhile, the time $t_{\rm inv}$ of the management period advances by one $\tau$ at a time, as shown below.
\begin{eqnarray}
	t_{\rm inv} &\in& \bm{t}_{\tau, {\rm inv}} \ ,\\
	\bm{t}_{\tau, {\rm inv}}
	&=& \{ t_{\rm inv}^{s}, t_{\rm inv}^{s}\!+\!\tau, t_{\rm inv}^{s}\!+\!2\tau, \cdots, t_{\rm inv}^{e} \} \ . 
\end{eqnarray}
Here, $t_{\rm inv}^{s}$ and $t_{\rm inv}^{e}$ represent the start and end times of the management period, respectively. 
By temporally shifting $\bm{t}_{\tau, {\rm inv}}$ by one,
$\tau$ types of unduplicated $\bm{t}_{\tau, {\rm inv}}$ can be constituted.
At each time of $t_{\rm test}$ and $t_{\rm inv}$, the autoencoder is relearned.

\subsection{Comparison with fine tuning}
In addition to the data augmentation in the preceding section, data utilization of multiple time scales by fine tuning can also be envisaged. 
Fine tuning refers to a machine learning method in which prior learning is undertaken with large-scale data, and fine tuning of the model is done with small-scale data of the target tasks thereafter.

When the data of multiple time scales is utilized and fine tuning is carried out, the learning data set that is used in prior learning $\bm{S}_{\tau^*}^{\rm pre}(t)$ is as follows for the time scale $\tau^*$ of the target tasks:
\begin{equation}
	\bm{S}_{\tau^*}^{\rm pre}(t) = \{ \bm{r}^{\dag}_\tau(t^\prime) \ | \ t^\prime \in \bm{t}_{\tau,{\rm train}}(t), \tau \in 1 \sim \tau^*-1 \} \ . 
	\label{eq:eqft1}
\end{equation}
Subsequently, the learning data set that is used for fine tuning of the model $\bm{S}_{\tau^*}^{\rm ft}(t)$ is as follows:
\begin{equation}
	\bm{S}_{\tau^*}^{\rm ft}(t)= \{ \bm{r}^{\dag}_\tau(t^\prime) \ | \ t^\prime \in \bm{t}_{\tau, {\rm train}}(t), \tau \in \tau^* \} \ . 
	\label{eq:eqft2}
\end{equation}
In this way, prior learning is conducted with the data of a shorter time scale than the time scale $\tau^*$ of the target tasks, and fine tuning of the model is done with the data of the time scale of the target tasks after that. 
For the purpose of comparison, the same hyperparameters as in Table \ref{tab:hyperparam} are applied.

\subsection{Generalization error}
To verify the effectiveness of data augmentation (hereinafter, MTS: multiple time scales) by Equation (\ref{eq:eqmts}), a comparison of the generalization error was conducted with the period from January 2016 to October 2017 as the test period. 
The RMSE of the test period was used as the evaluation index for generalization error:
\begin{equation}
	{\rm RMSE}_{\tau} = \displaystyle \sqrt{
	\frac{1}{|\bm{t}_{\tau, {\rm test}}|\cdot N}
	\sum_{t \in \bm{t}_{\tau, {\rm test}} }^{|\bm{t}_{\tau, {\rm test}}|} \sum_{i=1}^{N} \left( r_{i,\tau}(t) - \hat{r}_{i,\tau}(t) \right)^2 \ . 
	}
	\label{RMSE}
\end{equation}
As the comparison methods, a model in which learning is done by just a single time scale based on Equation (\ref{eq:eqsts}) (hereinafter, STS: single time scale)
and a model in which learning is done by fine tuning based on Equations (\ref{eq:eqft1}) and (\ref{eq:eqft2}) (hereinafter, FT: fine tuning) were employed. 
In addition, Table \ref{tab:optimum} shows the options accompanying optimization of the model, including the activation function of the middle layer. 
For this sub-section, the compression ratio $C$[\%] ($=$[Number of dimensions of the middle layer $M$/Number of dimensions of the input layer $N$] $\cdot 100\%$) of the middle layer was set at $50\%$.

\begin{table}[t]
	\centering
	\caption{Items to be Optimized}
	\label{tab:optimum}
	\begin{tabular}{c|c}
		\hline
		Item & Option \\
		\hline
		Fine tuning & Yes, no \\
		Activation function of the middle layer & Linear function, tanh function \\
		\hline
	\end{tabular}
\end{table}
\par

Figure \ref{fig:generr} shows the generalization error ${\rm RMSE}_{\tau^*}$ for the time scale $\tau^*$ of the target tasks in the test period. 
The insights obtained are listed below.
\begin{itemize}
	\item The generalization error for MTS and FT is smaller than that for STS. 
		Therefore, the combined use of the data of multiple time scales is significant in learning. 
		The reason for this may be because a multifactor model does not have a characteristic time scale; hence, the data of the time scales that differ from the target tasks can also contribute to learning.

	\item The generalization error for MTS is smaller than that for FT. 
		Therefore, it is better to simultaneously learn the data of multiple time scales than to apply fine tuning. 
		The reason for this may be because fine tuning is done with a small amount of learning data, which causes overlearning due to the limitation of the data sample.

	\item In MTS and FT, the generalization error of a linear function is smaller than that of a tanh function when it comes to the activation function of the middle layer. 
		This may involve overlearning as well. 
		In the simpler STS, a linear function is better at $\tau^* \leq 13$, but a tanh function is better at $\tau^* \geq 14$.

	\item In MTS and FT, the larger the time scale $\tau^*$, the smaller is the generalization error. 
		From this perspective, the effects of the combined use of multiple time scales can also be checked. 
		The larger the time scale $\tau^*$, the more the total amount of learning data is increased; hence, reduction of the generalization error can be expected.
\end{itemize}
From the above results, a model in which MTS (no fine tuning) is used for the learning method of the autoencoder and a liner function is used as the activation function of the middle layer is regarded as optimal, so this was employed in the experiment below.

\subsection{Compression ratio of the middle layer}
\begin{figure}[htbp]
	\centering
	\includegraphics[width=0.9\linewidth]{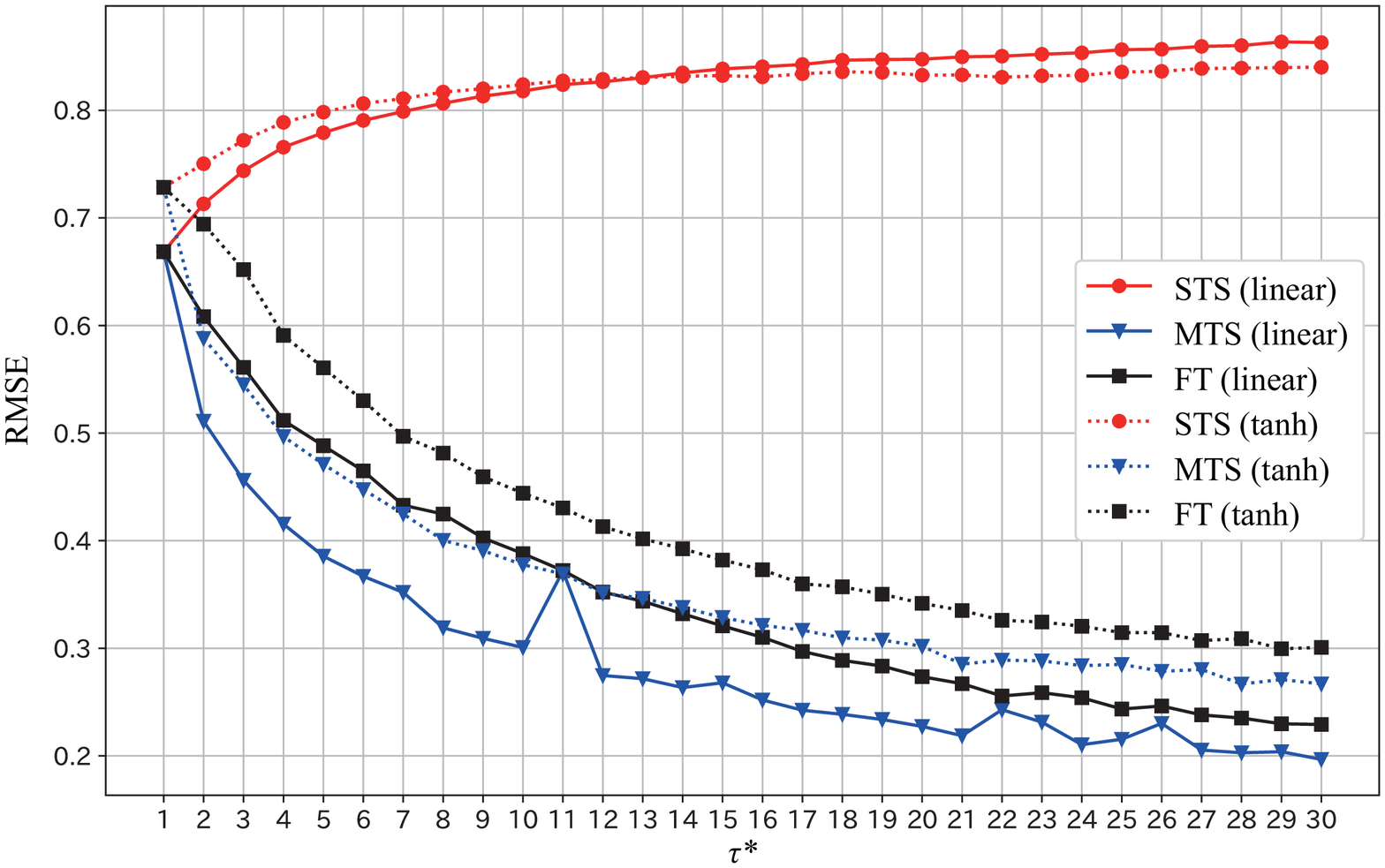}
	\caption{Generalization error (RMSE) of the test period in the time scale of the target tasks. The numbers in parentheses indicate the activation function of the middle layer.}
	\label{fig:generr}
\vspace{15mm}
	\centering
	\includegraphics[width=0.9\linewidth, height=50mm]{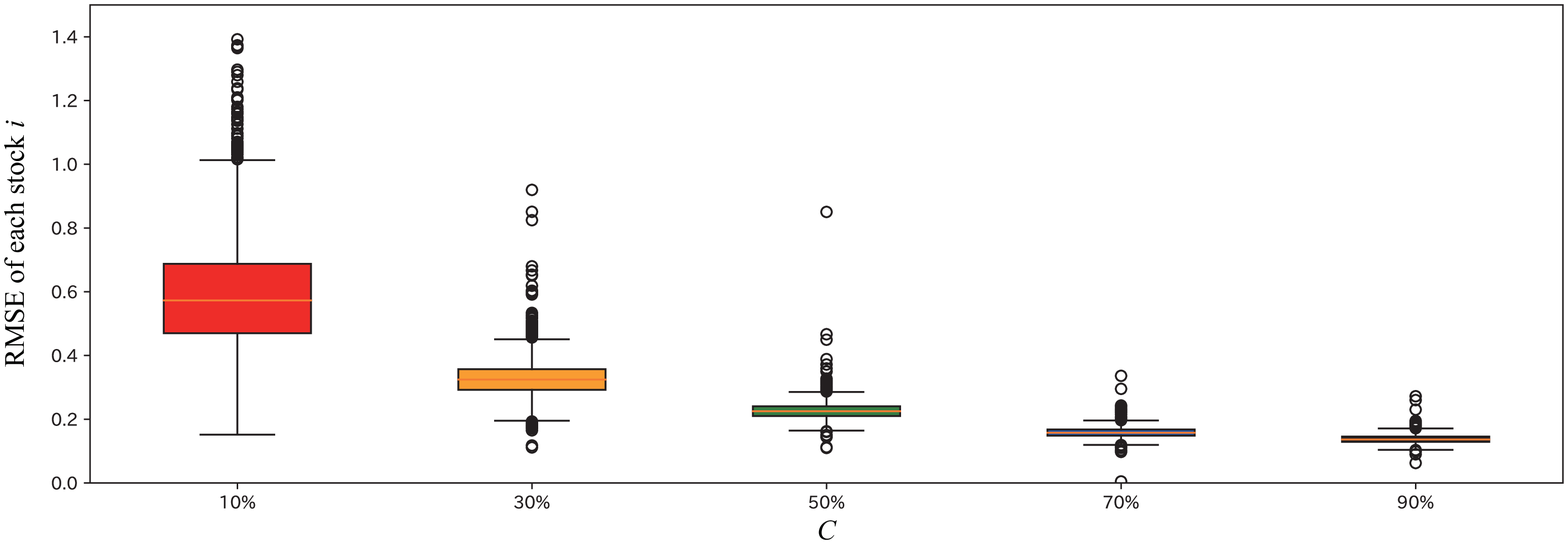}
	\caption{Relationship between compression ratio $C$ of the middle layer and generalization error (RMSE).}
	\label{fig:comp}
\end{figure}

Next, the compression ratio $C$ of the middle layer is optimized. 
Application to portfolio management is assumed and the time scale is set at $\tau^*=20$. 
Figure \ref{fig:comp} shows the generalization error ${\rm RMSE}_{\tau^*}$ in the event that the compression ratio $C$ is varied. 
The greater the compression ratio $C$ is, the easier it is to restore it, so the generalization error can be reduced, but the significance of performing dimensional compression with the autoencoder is lost. 
However, because the generalization error generally converges at a compression ratio of approximately $50\%$, $50\%$ is adopted as the optimal compression ratio $C$ in the experiment below.

\section{Management Simulation}

\subsection{Management method}
Portfolio management utilizing an autoencoder was undertaken as the practical application of the proposed method. 
In this section, the time scale is set at $\tau^*=20$, and the portfolio is rebalanced with a monthly scale.

The realized value of the return at time $t$ $\bm{r}_{\tau^*}^{\dag}(t)$ is inputted into a learned autoencoder, and the restored value $\bm{r}_{\tau^*}^{\dag}(t)$ is obtained. 
Based on Equation (\ref{eq7}) and $\sigma_{\tau^*}(t)$, the deviation from the realized value $\bm{r}_{\tau^*}^{\dag}(t)$ (i.e., abnormal return) becomes the following for each stock $i$:
\begin{equation}
	\epsilon_{i,\tau^*}(t) = r_{i,\tau^*}(t)-\hat{r}_{i,\tau^*}(t) \ .
\end{equation}
This corresponds to Equation (\ref{eq5}). 
However, because the degree of difficulty of learning of the autoencoder differs for each stock $i$,
the difference in the degree of difficulty of learning is corrected by dividing the restoration error $\xi_{i,\tau^*}(t)$ of the learning data, as shown below.
\begin{eqnarray}
	\epsilon^\dagger_{i,\tau^*}(t) &=& \frac{\epsilon_{i,\tau^*}(t)}{\xi_{i,\tau^*}(t)} \ ,\\ 
	\xi_{i,\tau^*}(t) &=& \displaystyle \sqrt{
	\frac{1}{|\bm{t}_{\tau^*, {\rm train}}(t)|}
	\sum_{t^\prime \in \bm{t}_{\tau^*, {\rm train}}(t)}^{|\bm{t}_{\tau^*, {\rm train}}(t)|}
	\left( r_{i,\tau^*}(t^\prime) - \hat{r}_{i,\tau^*}(t^\prime) \right)^2
	} \ .
	\label{eq:anm}
\end{eqnarray}
Each time $t$ advances, the learning data set $\bm{S}_{\tau^*}(t)$ is updated, and the autoencoder is relearned.

According to Equation (\ref{eq5}), the abnormal return $\epsilon^\dagger_{i,\tau^*}(t)$ is composed of the specific return $\eta_{i,\tau^*}(t)$ and mispricing $A_{i, \tau^*}(t)$. 
The specific return will not affect the stock price changes from $t$ as long as these are already incorporated results. 
However, because mispricing can become the cause of correction of the stock price from time $t$ onwards, we want to extract the stocks $i$ for which $A_{i, \tau^*}(t)$ is large both positively and negatively in portfolio management. 
However, because $\eta_{i,\tau^*}(t)$ cannot be observed, reference is made to $\epsilon^\dagger_{i,\tau^*}(t)$ as the proxy variable of $A_{i, \tau^*}(t)$, as noted in Sub-section 2.3.

Accordingly, the stocks were grouped into five quantiles in accordance with the size of the abnormal return $\epsilon^\dagger_{i,\tau^*}(t)$ at each time point, where the first quantile (1Q) is set as the minimum quantile and the fifth quantile (5Q) is set as the maximum quantile. 
As long as the model can detect abnormalities properly, the undervalued stocks are classified in 1Q and the overvalued stocks are classified in 5Q. 
The weighting is set at an equal weight in each quantile, and portfolio management with a cycle of $\tau^* (=20)$ is run.

\subsection{Tendency of stock selection}
First, in the test period from January 2016 to October 2017, the tendencies of the stocks to be selected were checked. 
In the case of a long-short portfolio that is long in undervalued stocks (1Q) and short in overvalued stocks (5Q),
when $\epsilon^\dagger_{i,\tau^*}(t)$ fluctuates positively and negatively, long and short are repeatedly back and forth for the same stock $i$, so the trading cycle is fast. 
As a result, the burden of operations and transaction costs increase. 
To check this tendency, the fluctuation of $\epsilon^\dagger_{i,\tau^*}(t)$ is checked by the following two methods.\\

\noindent
\underline{Method 1: Autocorrelation}
	\begin{enumerate}[(step 1)]
		\item The abnormal return $\epsilon^\dagger$ of all stocks is acquired in the test period.
		\item The autocorrelation of the abnormal return $\epsilon^\dagger$ is calculated for each stock $i$.
	\end{enumerate}

\noindent
\underline{Method 2: Jaccard coefficient}
	\begin{enumerate}[(step 1)]
		\item The selected stocks (stocks included in 1Q or 5Q) are acquired in the test period.
		\item The concordance rate with the time $t+n \cdot \tau^*$ is evaluated by the Jaccard coefficient for the selected stocks at each time $t$.
	\end{enumerate}
\begin{eqnarray}
\!\!\!\!A(n)&\!\!\!\!=\!\!\!\!&
				\frac{1}{| {\bm t}_{\tau, {\rm test}}|}
				\sum_{t \in {\bm t}_{\tau, {\rm test}}}^{|{\bm t}_{\tau, {\rm test}}|}
				\displaystyle\frac{|I_{\rm LS}(t) \cap I_{\rm LS}(t+n \cdot \tau^*)|}{|I_{LS}(t) \cup I_{\rm LS}(t+n \cdot \tau^*)|} \ , \\
\!\!\!\!B(n)&\!\!\!\!=\!\!\!\!&
				\frac{1}{| {\bm t}_{\tau, {\rm test}}|}
				\sum_{t \in {\bm t}_{\tau, {\rm test}}}^{|{\bm t}_{\tau, {\rm test}}|}
				\displaystyle\frac{|I_{\rm L}(t) \cap I_{\rm L}(t+n \cdot \tau^*)| + |I_{\rm S}(t) \cap I_{\rm S}(t+n \cdot \tau^*)|}{|I_{\rm LS}(t) \cup I_{\rm LS}(t+n \cdot \tau^*)|} \ , \\
\!\!\!\!C(n)&\!\!\!\!=\!\!\!\!&
				\frac{1}{| {\bm t}_{\tau, {\rm test}}|}
				\sum_{t \in {\bm t}_{\tau, {\rm test}}}^{|{\bm t}_{\tau, {\rm test}}|}
				\displaystyle\frac{|I_{\rm L}(t) \cap I_{\rm S}(t+n \cdot \tau^*)| + |I_{\rm S}(t) \cap I_{\rm L}(t+n \cdot \tau^*)|}{|I_{\rm LS}(t) \cup I_{\rm LS}(t+n \cdot \tau^*)|} \ .
\end{eqnarray}

Here, $I_{\rm LS}(t)$ is the stock set that is included in 1Q or 5Q at time $t$ (long-short portfolio), $I_{\rm L}(t)$ is the stock set (long portfolio) that is included in 1Q at time $t$,
and $I_{\rm S}(t)$ is the stock set (short portfolio) that is included in 5Q at time $t$. 
The time scale is set at $\tau^* = 20$. 
If $B(n)<C(n)$, it means that there are many stocks that fluctuate back and forth between 1Q and 5Q, and if $B(n)>C(n)$, it means that there are many stocks that remain permanently in 1Q or 5Q. 
$A(n)$ represents the sustainability of the stocks to be selected.

\begin{figure}[htbp]
	\begin{tabular}{cc}
		\begin{minipage}[t]{\hsize}
			\centering
			\includegraphics[width=0.9\linewidth]{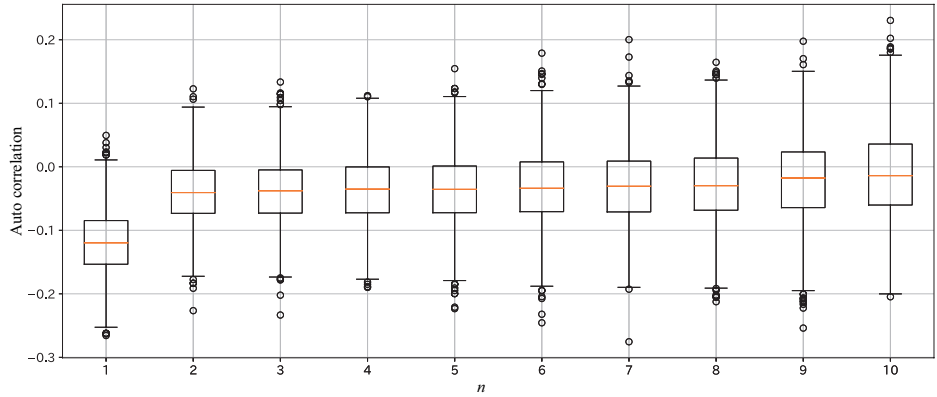}
			\subcaption{Autocorrelation (method 1). The $20$ kinds simulated by different start dates $t^s_{\rm inv}$ are averaged by each stock $i$, and the difference among stocks is shown in the box plot.}
		\vspace{5mm}
		\end{minipage}\\
		\begin{minipage}[t]{\hsize}
			\centering
			\includegraphics[width=0.9\linewidth]{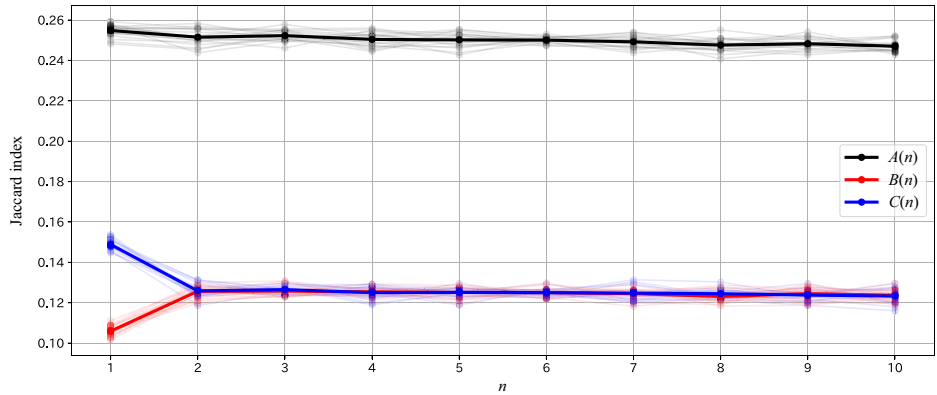}
			\subcaption{Jaccard coefficient (method 2). The $20$ kinds simulated by different start dates $t^s_{\rm inv}$ are indicated by the thin line, while the average value thereof is indicated by the thick line.}
		\end{minipage}
	\end{tabular}
	\caption{Situation where the degree of abnormality $A_{i, \tau^*}(t)$ fluctuates.}
	\label{fig:tendency}
\end{figure}

Figure \ref{fig:tendency} shows the results given by $\tau^*$ ($=20$) kinds of simulations
that were independently simulated with different start date $t^s_{\rm inv}$ of the management period.
From Figure \ref{fig:tendency}(a), the self-correlation of each stock generally exhibits negativity at $n=1$. 
In other words, the abnormal return $\epsilon^\dagger_{i,\tau^*}(t)$ has a tendency to fluctuate easily between positive and negative. 
Then, the autocorrelation weakens abruptly at $n \geq 2$ and becomes more or less uncorrelated. 
Moreover, from Figure \ref{fig:tendency}(b), the fluctuation between positive and negative of the abnormal return can be confirmed from $B(1)<C(1)$. 
It then becomes $B(n) \simeq C(n)$ at $n \geq 2$. 
Based on the above, the abnormal return of each stock fluctuates readily between positive and negative, so the number of stocks to be reverse traded increases in the rebalancing after time $\tau^*$ ($n=1$). 
To reduce these reversing trades, the following restrictions are introduced:
\begin{itemize}
	\item The stocks that were long in 1Q are outside the scope of application even if they are included in 5Q after time $\tau^*$ ($n=1$).
	\item The stocks that were short in 5Q are outside the scope of application even if they are included in 1Q after time $\tau^*$ ($n=1$).
\end{itemize}
From $A(n) \simeq 0.25$, the stocks for which the abnormal return is large positively and negatively are easily sustained. 
This is consistent with 
the volatility clustering. 
Moreover, from $B(n) \simeq C(n) \simeq 0.13$, approximately $13\%$ of the stocks are repeated
irrespective of whether they are held long or short.

\subsection{Management performance}
The period from January 2018 to December 2021 after the test period is set as the management period, and two kinds of management simulation are undertaken. 
First, the earned returns based on the long portfolios for each quantile, for which verification of the effectiveness of the stock selection based on the abnormal return $\epsilon^\dagger_{i,\tau^*}(t)$ (proxy variable of mispricing) is to be done, are compared. 
As the benchmarks at that time, the active return is evaluated by deducting the average return of all management target stocks. 
Second, actual portfolio management is assumed, and a long-short portfolio that is long in the underpriced stocks of 1Q and short in the overpriced stocks of 5Q is constructed. 
The effects of the restrictions for preventing the aforementioned reversing trades are verified based on the spread return thereof.

\begin{figure}[htbp]
	\begin{tabular}{cc}
		\begin{minipage}[t]{\hsize}
			\centering
			\includegraphics[width=0.9\linewidth]{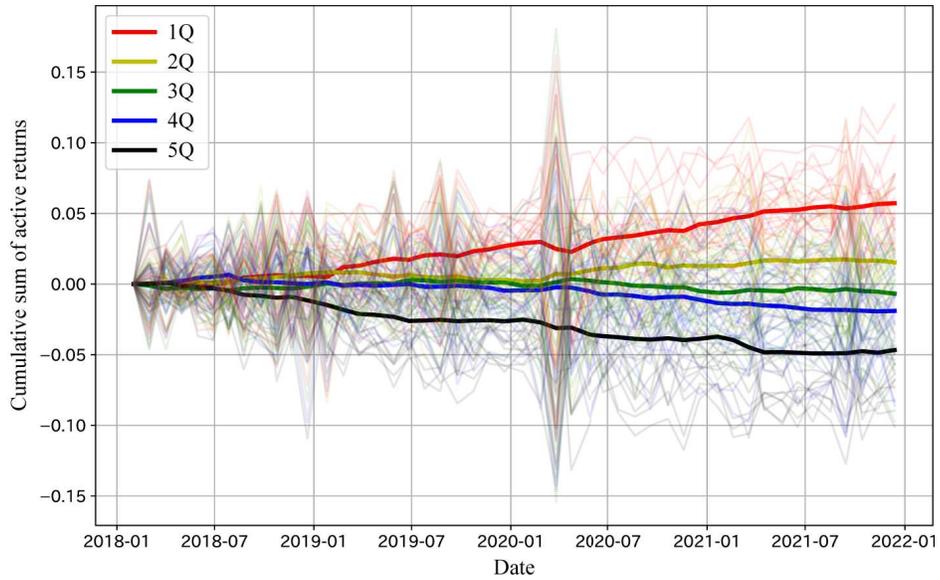}
			\subcaption{Cumulative sum of the active return based on a long portfolio by each quantile (case of no restrictions).}
			\vspace{5mm}
		\end{minipage}\\
		\begin{minipage}[t]{\hsize}
			\centering
			\includegraphics[width=0.9\linewidth]{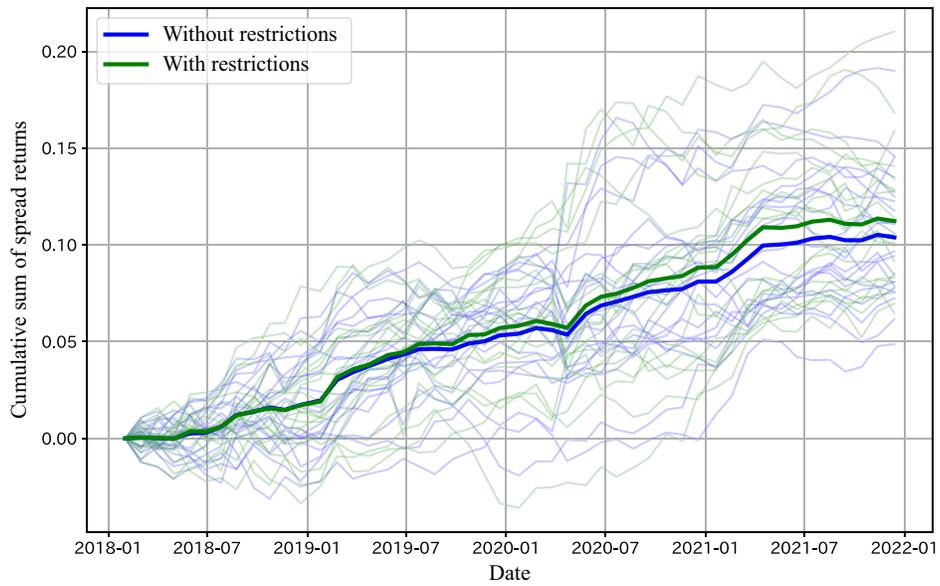}
			\subcaption{Cumulative sum of the spread return based on a long-short portfolio.}
		\end{minipage}
	\end{tabular}
	\caption{Management performance in the stock market. The $20$ kinds simulated by different start dates $t^s_{\rm inv}$
	are indicated by the thin line, while the equal weight portfolio (average value) thereof is indicated by the thick line.}
	\label{fig:運用パフォーマンス}
\end{figure}

Figure \ref{fig:運用パフォーマンス} shows the results. 
From Figure \ref{fig:運用パフォーマンス}(a), the active return is in quantile order, and the appropriateness of the stock selection based on the abnormal return $\epsilon^\dagger_{i,\tau^*}(t)$ can be confirmed. 
Moreover, from Figure \ref{fig:運用パフォーマンス}(b), the cumulative sum of the respective spread returns rises rightwards in the graph. 
In addition, the management performance can be stabilized by averaging all of these by an equal weight portfolio. 
Further, higher return is obtained by establishing the restrictions, and burdensome reversing trades can be alleviated.

\subsection{Correction of mispricing}
Finally, in the same management period (January 2018 to December 2021), the following cross section regression analysis was performed to directly verify whether or not correction pressure is acting on the mispricing:
\begin{equation}
	r_{i,\tau^*}(t+\tau^*)= \beta_{\tau^*}(t) \cdot \epsilon^\dagger_{i,\tau^*}(t) + \kappa_{\tau^*}(t) + \xi_{i,\tau^*}(t) \ .
\end{equation}
The time scale is set at $\tau^*=20$, as in the portfolio management in the preceding sub-section, and the regression coefficient $\beta_{\tau^*}(t)$ is evaluated. 
$\kappa_{\tau^*}(t)$ is the intercept, and $\xi_{i,\tau^*}(t)$ is the residue.

\begin{figure}[t]
	\centering
	\includegraphics[width=0.9\linewidth]{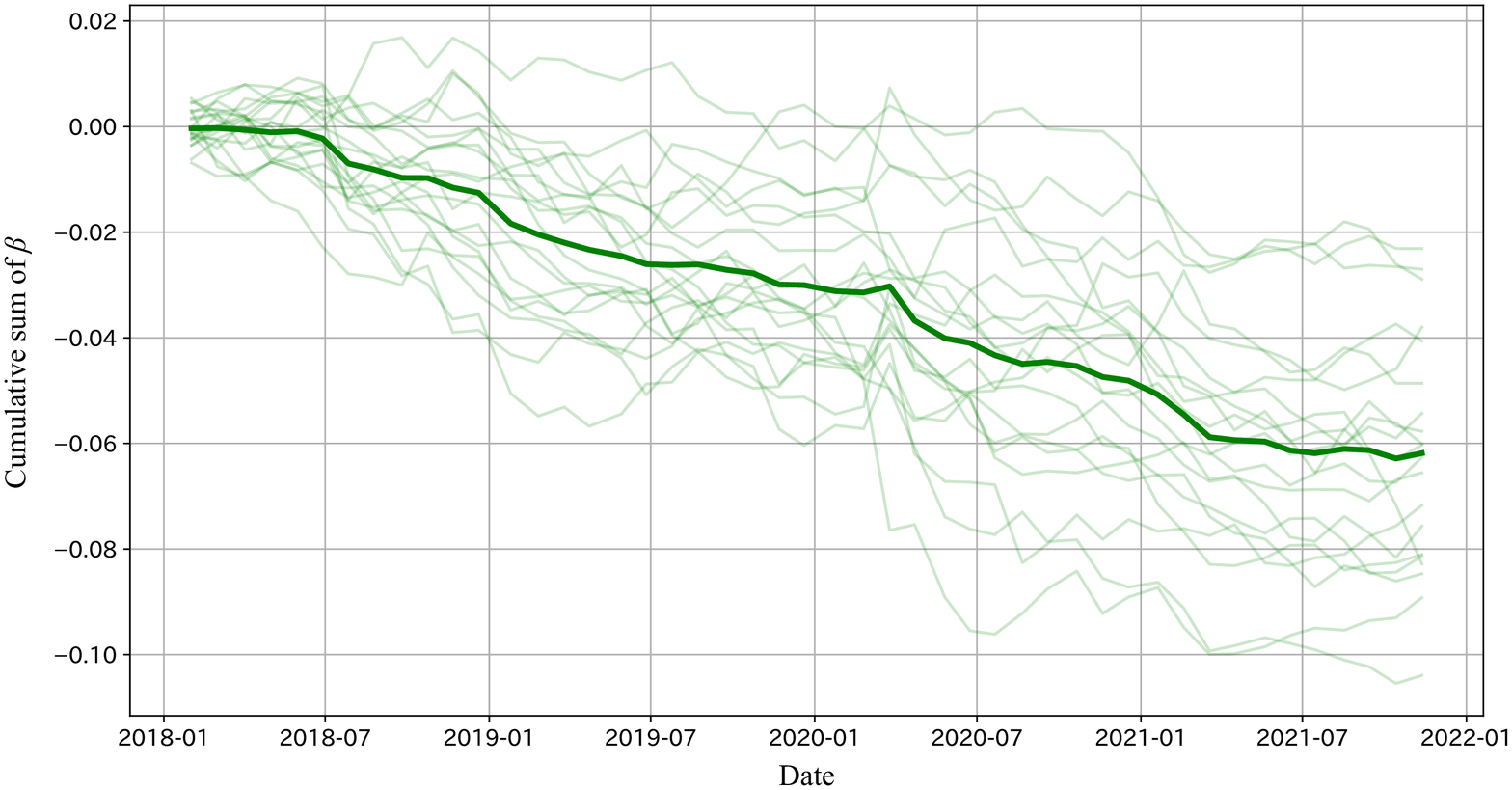}
	\caption{Grounds whereby mispricing is corrected.
	The cumulative sum of $\beta_{\tau^*}(t)$ of the $20$ kinds 
	simulated by different start dates $t^s_{\rm inv}$
	are indicated by the thin line, while the average value thereof is indicated by the thick line.}
	\label{fig:repair}
\end{figure}

Figure \ref{fig:repair} shows the results. 
The cumulative sum of the regression coefficient $\beta_{\tau^*}(t)$ is stably trending in the negative direction, and $r_{i,\tau^*}(t+\tau^*)$ easily moves in the direction of correcting the abnormal return $\epsilon^\dagger_{i,\tau^*}(t)$. 
Therefore, the abnormal return $\epsilon^\dagger_{i,\tau^*}(t)$ can be utilized as the proxy variable of the mispricing $A_{i,\tau^*}(t)$.

\section{Application to the FX Market}

\subsection{Dataset}
As the autoencoder is an augmentation of the multifactor model, it can be applied universally provided that the return data are observable. 
In fact, the existence of common factors like carry, value, and trend has been reported chiefly in the FX market\cite{FX1,FX2,FX3,FX_factor}.
Accordingly, the proposed method verifies whether or not it is effective in not only the stock market but also the FX market. 
The currencies of the G10 member countries
are used as the management targets,
and specifically a total combination of Japanese yen, US dollars, the Euro, the British pound, the Canadian dollar, the Swiss franc, and the Swedish krona ($21$ currency pairs) is employed. 
The period from August 2010 to December 2021 was used as the test period,
of which the generalization performance of the model was checked for a test period from January 2016 to October 2017,
and simulation was performed for a management period from January 2018 to December 2021.

\subsection{Generalization error}
In the test period from January 2016 to October 2017, the generalization error (RMSE) in Equation (\ref{RMSE}) is calculated. 
As in Sub-section 4.3, MTS, FT, and STS are employed, and the generalization error is compared with six models for cases in which the middle layer of each autoencoder is set to a linear function or tanh function.

\begin{figure}[htbp]
	\centering
	\includegraphics[width=0.9\linewidth]{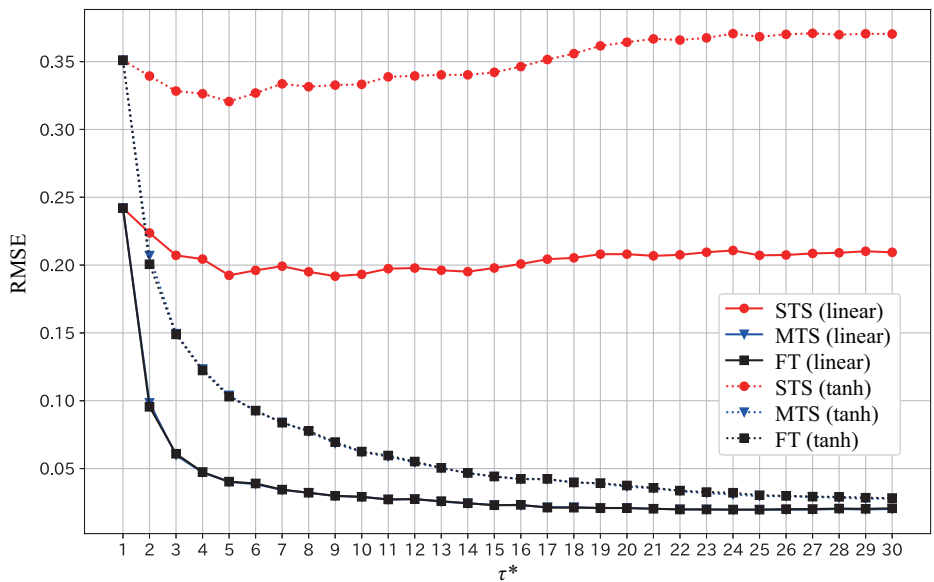}
	\caption{Generalization error (RMSE) of the test period in the FX market.
	The items in parentheses indicate the activation function of the middle layer.}
	\label{fig:FXgenerr}
\end{figure}

Figure \ref{fig:FXgenerr} shows the results. 
The generalization error was reduced by utilizing the data of multiple time scales in the FX market as well. 
As for the differences with the results of the stock market, the generalization error of FT is equivalent to that of MTS. 
The reason for this may be that the complexity of the model is reduced by the decrease in the number of stocks (i.e., the dimension of the input layer),
and overlearning due to fine tuning tends not to occur. 
However, as there is no characteristic time scale in the original factor model, it is believed that the generalization error can be reduced in the same manner as MTS.

\subsection{Management method}
The same portfolio management as that presented in Section 5 is performed for the FX market as well. 
However, because the FX market differs from the stock market from two perspectives, the management method is altered by taking these into consideration. 
The first difference is that the liquidity of the FX market is higher than that of the stock market so the correction speed for mispricing may be higher. 
Accordingly, although the time scale is set at $\tau^*=20$ in the case of the stock market, management with a three-day cycle as $\tau^*=3$ is performed in the case of the FX market.

The second difference is that while management was carried out with approximately $1,700$ stocks in the stock market, management is done by $21$ currency pairs in the FX market. 
It is also possible to increase the number of pairs by employing currencies of other than the G10 member countries,
but minor currencies are included. 
As the number of trading participants is small for minor currencies, the liquidity is low, and trades may not be successfully concluded in actual management. 
Moreover, because the volatility of currencies tends to be large,
this may become an obstacle to learning.
Therefore, in this study, minor currencies fall outside the scope of application, and the 21 currency pair management shown in Sub-section 6.1 was conducted. 
However, when a five quantile portfolio is constructed in the same manner as the stock market, only four currency pairs are incorporated in each portfolio, so risk cannot be spread adequately. 
Accordingly, in the management of the FX market, the portfolio is built by setting the number of quantiles as two. 
If the model is functioning properly, the underpriced currency pairs will be classified in the first quantile (1Q) and overpriced currency pairs will be classified in the second quantile (2Q).

\subsection{Management performance}
\begin{figure}[htbp]
	\centering
	\includegraphics[width=0.9\linewidth]{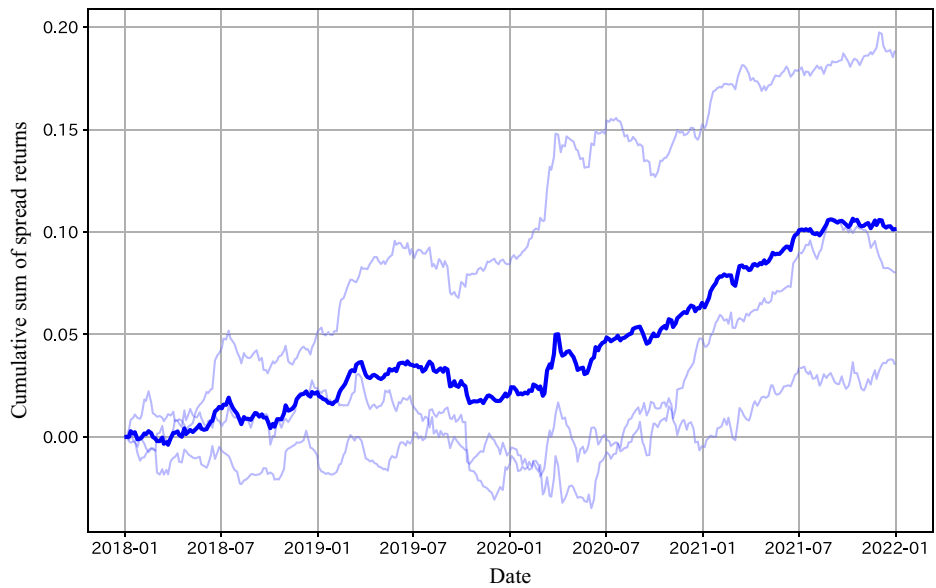}
	\caption{Cumulative return of the spread return of the long-short portfolio in the FX market.
	The three kinds simulated by different start dates $t^s_{\rm inv}$
	are indicated by the thin line, whereas the equal weight portfolio (average value) thereof is indicated by the thick line.}
	\label{fig:FXspread}
\end{figure}

In the management period from January 2018 to December 2021,
a long-short portfolio that takes long position in the first quantile (1Q) and short position in the second quantile (2Q) without the restrictions mentioned in Section 5.2 was implemented.
At that time, three kinds of start date $t^s_{\rm inv}$ in the management period
can be set, so these were simulated independently, and Figure \ref{fig:FXspread} shows the results of these. 
The cumulative sum of each spread return rises rightwards in the graph.
In addition, the management performance can be stabilized by averaging all values by an equal weight portfolio. 
This suggests that the data augmentation and portfolio management in this study operate not only in the stock market but also in markets more universally.

\section{Conclusion}
Data augmentation was conducted by the combined use of data of multiple time scales as a general solution for the problem of insufficient data, which is an issue in medium to long term asset management business. 
As one example of this, the focus herein was placed on a multifactor model that does not rely on characteristic time scales; the multifactors were extracted by the machine leaning of an autoencoder. 
Subsequently,  data augmentation was confirmed to reduce the generalization error more than in the case in which machine learning is performed with a single time scale. 
Next, as a practical application, the aforementioned autoencoder was utilized in portfolio management. 
Deviation from the theoretical value estimated with the autoencoder was regarded as mispricing, stock selection that used the correction pressure to a fair price was carried out, and the appropriateness of the stock selection was confirmed through a management simulation. 
Finally, application to the FX market was performed to confirm the general utility of this proposed method. 
As a result, the effectiveness of portfolio management using data augmentation and mispricing by multiple time scales was confirmed, as in the case of the stock market.

\section*{Acknowledgemet}
The authors would like to thank Mitsunori Hokao and Riku Tanaka of Daiwa Asset Management Co.Ltd. for useful discussions.
This research was partially supported by the Grant-in-Aid for Scientific Research (C) (20K11969)
from the Ministry of Education, Culture, Sports, Science and Technology of Japan.
The contents of this article are the personal views of its authors
and not the official views of the institutions with which they are affiliated. 

\end{document}